\documentclass[twocolumn,secnumarabic,amssymb, nobibnotes, aps, pre, superscriptaddress]{revtex4-1}

\usepackage{float}
\usepackage{graphicx}
\usepackage{amsmath,amssymb} 
\usepackage{epstopdf}
\usepackage{setspace}
\usepackage[shortlabels]{enumitem}
\usepackage{mathrsfs}
\usepackage{color,soul}
\usepackage[dvipsnames]{xcolor}
\usepackage{dcolumn}
\usepackage{comment}
\usepackage{upgreek}
\usepackage{bm}
\usepackage{svg}

\usepackage{hyperref}
\hypersetup{colorlinks = true,citecolor = blue,urlcolor = black}

\newcommand{\eq}[1]{Eq.~\eqref{#1}}

\newcommand{\fig}[1]{Fig.~\ref{#1}}

\newcommand{\movie}[1]{Movie~\ref{#1}}
\newcommand{\movies}[2]{Movies~\ref{#1}-\ref{#2}}

\newcommand{\appndx}[1]{SI~\ref{#1}}

\newcommand{\figsi}[1]{Fig.~SI.~\ref{#1}}

\newcommand{\abs}[1]{\left| #1 \right|}
\renewcommand{\vec}[1]{\boldsymbol{\mathbf{#1}}}
\newcommand{\tens}[1]{\boldsymbol{\mathbf{#1}}}
\newcommand{\del}{\vec{\nabla}}

\newcommand{\RE}[1]{\text{Re}\left[#1\right]}
\newcommand{\IM}[1]{\text{Im}\left[#1\right]}

\newcommand{\ps}{\abs{\psi}}
\newcommand{\app}{\ps^2}

\newcounter{siequation}
\newcounter{sifigure}
\newcommand{\dummyfig}[1]{\refstepcounter{sifigure}\label{#1}}
\newcounter{sisection}
\newcommand{\dummyapp}[1]{\refstepcounter{sisection}\label{#1}}
\newcounter{simovie}
\newcommand{\dummymov}[1]{\refstepcounter{simovie}\label{#1}}

\begin{document}
\title{Complex-tensor theory of simple smectics}

\author{Jack Paget}
\affiliation{Interdisciplinary Centre for Mathematical Modelling and Department of Mathematical Sciences, Loughborough University, Loughborough, Leicestershire LE11 3TU, UK.}
\author{Marco G. Mazza}
\affiliation{Interdisciplinary Centre for Mathematical Modelling and Department of Mathematical Sciences, Loughborough University, Loughborough, Leicestershire LE11 3TU, UK.}
\affiliation{Max Planck Institute for Dynamics and Self-Organization (MPIDS), Am Fa{\ss}berg 17, D-37077 G\"{o}ttingen, Germany.}
\author{Andew J. Archer}
\affiliation{Interdisciplinary Centre for Mathematical Modelling and Department of Mathematical Sciences, Loughborough University, Loughborough, Leicestershire LE11 3TU, UK.}
\author{Tyler N. Shendruk}
\email{t.shendruk@ed.ac.uk}
\affiliation{School of Physics and Astronomy, The University of Edinburgh, Peter Guthrie Tait Road, Edinburgh, EH9 3FD, UK.}

\begin{abstract}
{Smectic materials represent a unique state between fluids and solids, characterized by orientational and partial positional order, making them notoriously difficult to model, particularly in confining geometries. We propose a complex order parameter tensor to describe the local degree of lamellar ordering, layer displacement and orientation. The theory accounts for both dislocations and disclinations, as well as arrested configurations and colloid-induced local ordering. It considerably simplifies numerics, facilitating studies on the dynamics of topologically complex lamellar systems.}
\end{abstract}

\maketitle


The very properties that make smectic phases so interesting contrive to make them challenging to model. 
They are lamellar liquid crystals---stacking along one direction while maintaining liquid-like positional disorder within layers. 
By breaking translational symmetry, smectic layering allows dislocation defects\cite{harrison2000,Aharoni2017,repula2018}, while broken rotational symmetry of the layer normal allows disclinations\cite{harrison2000,suh2019}.
This makes smectics excellent systems for exploring self-assembly\cite{Hur2015,kim2019,rottler2020} and topology\cite{santangelo2005,matsumoto2015,zhang2020}, especially in confining geometries\cite{Yoon2010,kim2015,preusse2020} or in contact with micropatterned structures\cite{kim2018}.
Recent studies of confined smectic colloidal liquid crystals\cite{wittmann2021,Monderkamp2021} and defect annihilation in block copolymer films\cite{hur2018,Schneider2021} motivate the need for alternative theoretical descriptions that allow simulations to tackle more topologically complex geometries without relying on microscale models. 
Here, we propose a novel formalism to model simple lamellar smectics. 

Traditionally, lamellar ordering of smectics is described by expanding the mesogen density at each point $\vec{r}$, as $\rho\left(\vec{r},t\right) = \sum_{m=-\infty}^\infty \psi_m e^{i m \vec{q}\cdot\vec{r}} \approx \rho_0 + 2\RE{\Psi}$, where $\rho_0$ is the mean density and $\Psi=\ps e^{i(\vec{q}_0\cdot\vec{r}+\phi)}$. The argument of the exponential includes the wave vector $\vec{q}_0$ and an arbitrary phase $\phi$. 
Rearranging as $\Psi=\psi e^{i\vec{q}_0\cdot\vec{r}}$ allows one to write  the complex order parameter $\psi\left(\vec{r},t\right)=\ps e^{i\phi}$
in analogy to the order parameters for superfluids or superconductors\cite{degennes1972,Lubensky1990,navailles2009,Kamien2016,Zappone2020}. 
Commonly employed in Landau free energy expansions\cite{mukherjee2001,mukherjee2013}, $\ps = \left(\psi \psi^*\right)^{1/2}$ quantifies the extent of layering, while the phase $\phi \equiv \vec{q}_0\cdot\vec{u}$ encodes the layer displacement field $u\left(\vec{r}, t\right)$. 
Variation of $\phi\left(\vec{r}, t\right)$ indicates lamellar compression/dilation deformations (hereafter referred to jointly as compression). 
Though elegant and economical, this formalism has known shortcomings\cite{Pevnyi2014}. 
Fundamentally, {only $\RE{\Psi}$ is physical, and so} $\phi$ is not truly a single-valued function of position and $\psi$ is not an element of the unit circle $S^1$ but rather the orbifold $S^1 / \mathbb{Z}_2$\cite{chen2009,alexander2010,alexander2012}. 
As a result, the order parameter does not faithfully reflect the nematic symmetry  of the layer normal $\vec{N}$. 
Instead, the layer normal must be defined as a vector via the gradients $\del\Phi/\abs{\del\Phi}$, highlighting the relationship $\vec{q}_0=q_0\vec{N}$. 

In contrast to smectics, nematic theory uses the tensor $\tens{Q}=S\left(\vec{n} \otimes \vec{n}-\tens{\delta}/d\right)$ to collect both the scalar order parameter $S$ and apolar director $\vec{n}$ into a single order parameter, for dimensionality $d$ and identity matrix $\tens{\delta}$\cite{degennes1971}. 
The nematic order parameter simultaneously describes the extent of phase ordering and local direction of broken symmetry in an arbitrary reference frame. 
Thus, both the bulk and deformation free energy densities can written in terms of $\tens{Q}\left(\vec{r},t\right)$. 
Practically, $\tens{Q}$ enables numerical simulations of confined nematics\cite{emersic2019,duzgun2021}, colloidal liquid crystals\cite{hashemi2017,yuan2018,villada2021} and active fluids\cite{duclos2020,zhou2021,thijssen2021}, by treating defects as locally disordered cores, rather than singularities. 
For smectic-A liquid crystals, $\tens{Q}$-theories can be coupled to models of smectics\cite{ball2015,xia2021}. 

In this letter, we propose a tensorial order parameter field for lamellar smectics. 
The tensor $\tens{E}\left(\vec{r},t\right)$ is complex, symmetric, traceless and globally gauge invariant. 
It incorporates the extent of layering and relative layer displacement, previously described by $\psi$, as well as the layer normal orientation $\vec{N}$. 
It encompasses the advantages $\tens{Q}$-tensor formalism provides to nematics but for smectics. 
Here, we exclusively consider the simplest smectics, with only lamellar broken translational symmetry and layer-normal broken rotational symmetry. 
In liquid crystalline smectics, nematic director distortions are also possible, though assuming that twist and bend are prohibited is common. 
We focus solely on lamellae to show the suitability of $\tens{E}$ for resolving the phase ambiguity 
and demonstrate its utility in simulating confining geometries. 

\begin{figure}[tb]
    \centering
    \includegraphics[width=0.495\textwidth]{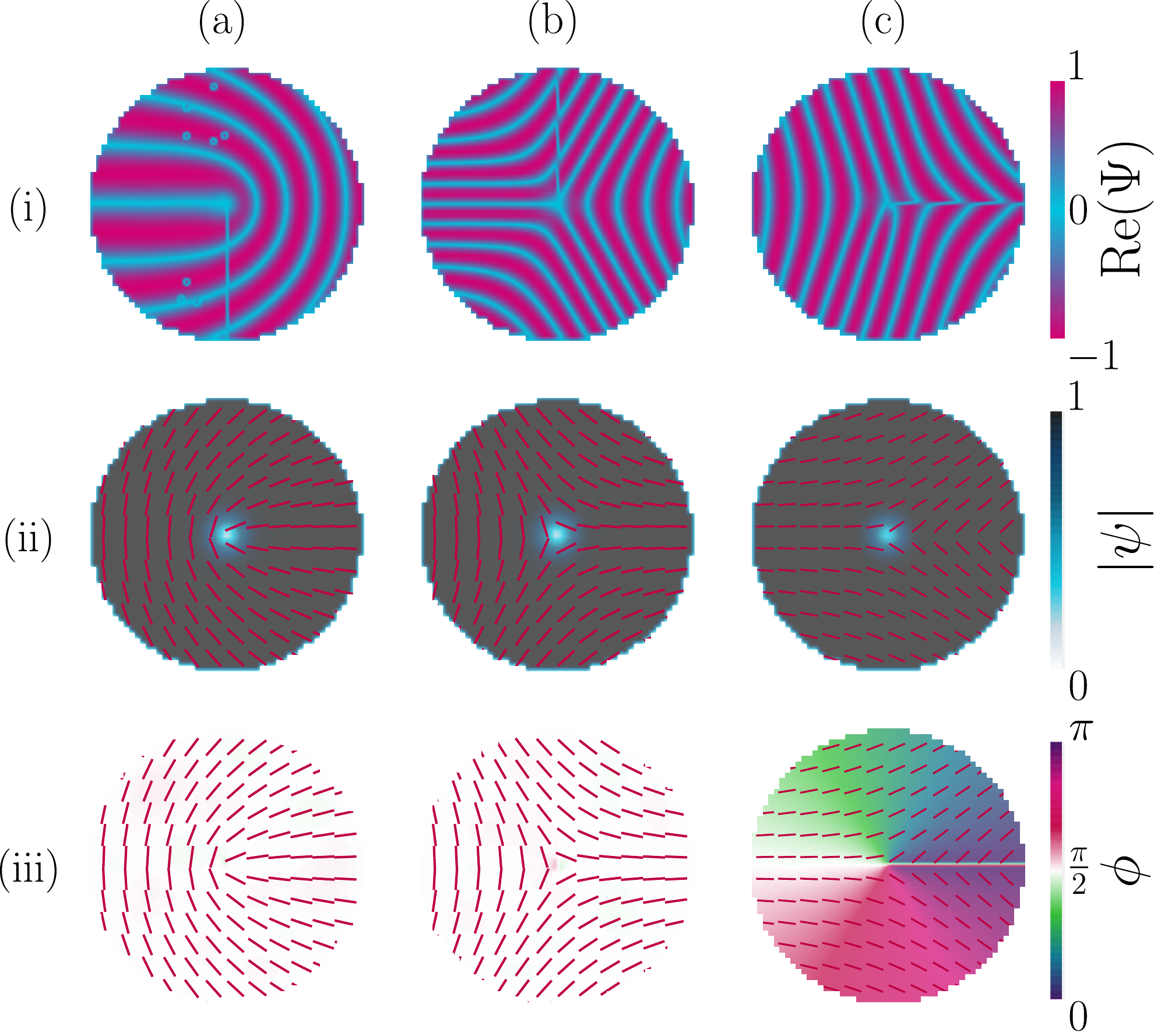}
    \caption{\small
        Simulations for $A=-1$ (lamellar state), $C=2$ and $\kappa^2=0.75$ in circular domains with boundary conditions requiring single defects. 
        \textbf{Columns} present three defect types: 
        \textbf{(a)} $+1/2$ disclination; 
        \textbf{(b)} $-1/2$ disclination; 
        \textbf{(c)} Edge dislocation. 
        \textbf{Rows} show plots of: 
        \textbf{(i)} $\RE{\Psi}$;
        \textbf{(ii)} $\ps$ with $\vec{N}$ overlayed;
        \textbf{(iii)} $\phi$ with $\vec{N}$. 
    }
    \label{fig:defects}
\end{figure}

To account for the apolar layer normal and resolve the phase ambiguity\cite{Pevnyi2014,zhang2020}, the smectic tensorial order parameter $\tens{E}\left(\vec{r},t\right)$ must contain the dyadic square of $\vec{N}$, making $\tens{E}$ symmetric. 
Furthermore, the absence of preferential directions within planar layers indicates local rotations about $\vec{N}$ are arbitrary. 
A traceless order parameter ensures linear terms do not contribute to the bulk free energy. 
Based on these considerations, we propose the complex-tensorial smectic order parameter
\begin{align}
 \label{eq:tensE}
 \tens{E}\left(\vec{r},t\right) &= \psi\left( \vec{N}\otimes \vec{N}-\frac{\tens{\delta}}{d} \right).  
\end{align}
The scalar order parameter $\psi\left(\vec{r},t\right) = \ps e^{i\phi}\in \mathbb{C}$ is the eigenvalue of $\tens{E}$ and the layer normal $\vec{N}\left(\vec{r},t\right) \in \mathbb{R}^d$ is the associated eigenvector. 
The order parameter is symmetric, traceless and globally gauge invariant (under $\tens{E}\to e^{i\theta}\tens{E}$ for arbitrary $\theta$); furthermore, it allows both $\vec{N}\to-\vec{N}$ and resolves the double-valued nature of $\phi$ through degeneracy of the eigenvalues (see Supplementary Information \appndx{sctn:methods}). 
The $d+1$ degrees of freedom embedded in $\tens{E}$ represent the extent of layering, layer displacement, and the unit vector. 

Though smectics and other lamellae have been modelled from many perspectives\cite{emelyanenko2015,alageshan2017,rottler2020,Schneider2021}, we consider a Landau free energy expansion. 
The total free energy density $f$ is the sum of bulk and two deformation (compression and curvature) terms. 
All contributions to the free energy must be real and invariant under $\tens{E}\to\tens{E}^*$, requiring pairings of $\tens{E}$ and its complex conjugate $\tens{E}^*$. 

\noindent\underline{\textit{Bulk}}:  
Since $\tens{E}$ is traceless, the bulk smectic free energy density can be written
\begin{align}
 \label{eq:bulkSm}
 f^\text{bulk} &= \frac{A}{2}E_{ij}E_{ji}^* + \frac{C}{4}\left(E_{ij}E_{ij}^*\right)^2+\ldots
\end{align}
where $C>0$, and Einstein summation convention is adopted.
When $A<0$  lamellar order is established, but when $A>0$ the fluid is isotropic. 
The bulk free energy does not depend on phase or layer normal, but only on $\ps$.
By \eq{eq:tensE}, $f^\text{bulk} = A\varrho^2\app/2 + C\varrho^4\ps^4/4$, where $\varrho=\left(d-1\right)/d$, which demonstrates the consistency between this complex tensor theory approach and scalar-based bulk free energies\cite{renn1988}. 
In the mean-field limit, \eq{eq:bulkSm} predicts a second order phase transition. 

\noindent\underline{\textit{Compression}}:
Lamellae possess two deformation modes: (i) compression, and (ii) curvature of the layers. 
We consider first compression free energies, which involve real derivatives of the tensor order parameter. 
The simplest such term is $E_{ij,k} E_{ij,k}^*$, where $k$ denotes the Cartesian direction of the gradient. 
Additional real terms could be constructed through combinations of similar forms, which would allow different deformation modes to possess differing elastic modulii. 
For clarity, we make a one-constant approximation
\begin{align}
 \label{eq:compression}
 f^\text{comp} &= b_1 E_{ij,k} E_{ij,k}^*,  
\end{align}
where $b_1$ is a layer compression elastic constant. 

\noindent\underline{\textit{Curvature}}:
Distortions from uniformly aligned layers come with a free energy cost, akin to a membrane curvature free energy density. 
We again make a one-constant approximation and keep only the simplest term
\begin{align}
 \label{eq:curv}
 f^\text{curv} &= b_2 E_{ij,kk} E_{ij,\ell\ell}^*, 
\end{align}
where $b_2$ is a bending modulus. 

\begin{figure*}[tb]
    \centering
    \includegraphics[width=0.95\textwidth]{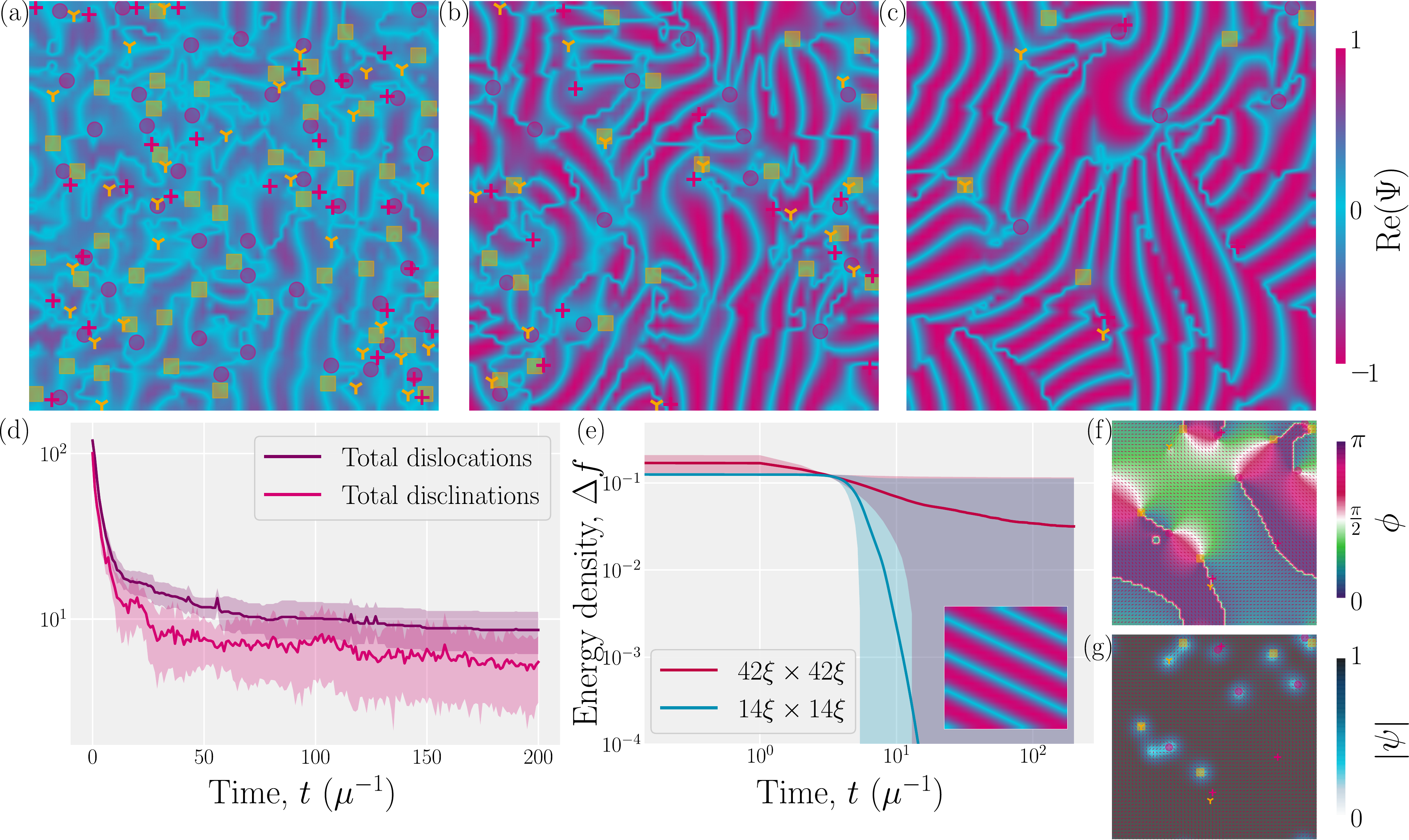}
    \caption{\small
    Deep quenched simple smectic, initialized from isotropic state ({$\ps\simeq0$ and random $\vec{N}$ and $\phi$}) with $A=-1$ (lamellar state), $C=2$, $\kappa^2=0.5$ and periodic boundary conditions. 
    \textbf{(a-c)} Snapshots of $\RE{\Psi}$ in a system size $42\zeta \times 42\zeta$ at times \textbf{(a)}$t=2\mu^{-1}$; \textbf{(b)} $5\mu^{-1}$; and \textbf{(c)} long-time limit of a kinetically arrested state.
    Pink crosses (yellow trilaterals) mark $+1/2$ ($-1/2$) disclinations. 
    Edge dislocations with winding number $\pm1$ denoted by pink circles and yellow squares. 
    \textbf{(d)} Average defect density. 
    \textbf{(e)} Free energy density, $\Delta f=f-f_{eq}$, for small ($14\zeta \times 14\zeta$) and large ($42\zeta \times 42\zeta$) systems. 
    \textbf{(e- inset)} Steady state for $14\zeta \times 14\zeta$.
    \textbf{(f)} Snapshot of $|\psi(\vec{r};t)|$ corresponding to \textbf{(c)}.
    \textbf{(g)} The corresponding $\phi(\vec{r},t)$ field. 
    }
    \label{fig:relax}
\end{figure*}

When the lamellae phase is free of deformations, minimizing the free energy produces the equilibrium values
\begin{align}
 \label{eq:equilibrium}
 \ps^\text{eq} = \sqrt{-\frac{A'}{C\varrho}}
 \quad &; \quad 
 q_0^\text{eq} = \sqrt{  \frac{b_1}{2b_2} }, 
\end{align}
where $A' \equiv A - 2b_1 q_0^2 + 2b_2 q_0^4$, in agreement with complex scalar Landau models\cite{mukherjee2001,mukherjee2013}. 
In this model, $\tens{E}$ is a hydrodynamic-scale field that does not involve layer spacing so identifying the wavenumber requires that covariant derivatives replace gradients\cite{chen1976,renn1988,lukyanchuk1998}. 
At equilibrium, the free energy is uniformly $f^\text{eq} = -\tfrac{\varrho}{2} \tfrac{A}{C} A' \left(1-\tfrac{\varrho}{2}\tfrac{A'}{A}\right)$. 
In addition to the layer thickness $2\pi/q_0^\text{eq}$, the free energy admits two length scales: (i) coherence length $\xi=\sqrt{b_1 /A}$ and (ii) penetration depth $\lambda=\sqrt{b_2/b_1}$. 
The coherence length $\xi$ characterizes the defect core size and the ratio of $\kappa=\lambda/\xi$ is a Ginzburg parameter. 
As in superconductors, $\kappa<1/\sqrt{2}$ is a type-I system, while $\kappa>1/\sqrt{2}$ is type-II\cite{degennes1972}. 
We also take the strong anchoring limit by fixing $\tens{E}$ at solid surfaces (limit of zero de Gennes-Kleman extrapolation length). 

Proven numerical schemes exist for minimizing the free energy of real $\tens{Q}$ tensors.
The numerical difficulty lies in extending the methodology to allow for complex tensor elements. 
We employ a gradient descent time evolution of $\tens{E}\left(\vec{r},t\right)$ in 2D {(see \appndx{sctn:methods}\cite{Liarte2015})}. 
Defining the total free energy 
$F = \int f dV$, we adopt a time-dependent Ginzburg-Landau model
\begin{align}
    \label{eq:tdGL}
    \mu \frac{\partial E_{\alpha\beta}}{\partial t} &= -\frac{\delta F}{\delta E_{\alpha\beta}^*} + \Lambda_{\alpha\beta},    
\end{align}
where $\mu$ is a mobility coefficient and $\tens{\Lambda}$ constrains $\tens{E}$ to be traceless and normal (see \appndx{sctn:lagrange}). 
It should be stressed that $\tens{E}\left(\vec{r},t\right)$ is the sole subject of all calculations---the complex amplitude $\psi\left(\vec{r}, t\right)$ and layer normal $\vec{N}\left(\vec{r}, t\right)$ are only found \textit{ex post facto}. 
Both $\ps$ and $\phi$ are calculated directly from contractions of $\tens{E}$ with itself, while $\vec{N}$ is found via eigen-decomposition (see \appndx{sctn:methods}). 
Defects are identified from the $\vec{N}$ and $\phi$ fields (see \appndx{sctn:defects}). 
While our approach circumvents the ambiguity of $\psi$ as a double-valued function $\RE{\psi} \pm i\IM{\psi}$\cite{Pevnyi2014,zhang2020}, visualizations and post-processed layer displacements do not. 
In particular, $\RE{\Psi}$ explicitly reveals lamellar structure (see \fig{fig:defects} and \appndx{sctn:viz}); however, these possess aberrations due to the phase ambiguity that $\tens{E}$ itself does not.  


\begin{figure*}[tb]
    \centering
    \includegraphics[width=0.95\textwidth]{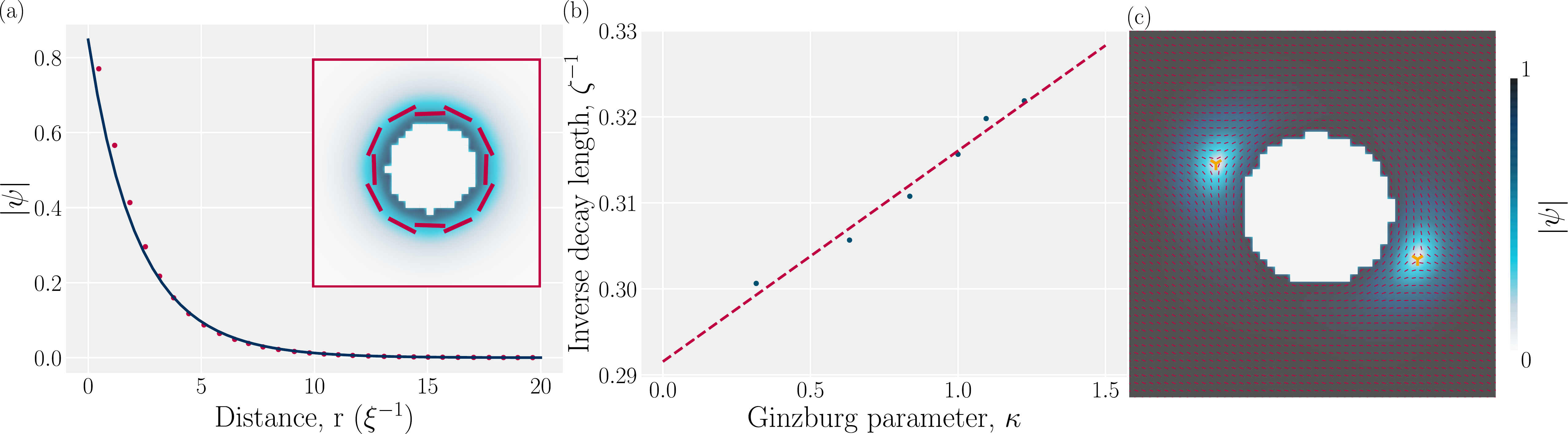}
    \caption{\small
    \textbf{(a)} Circular inclusion embedded in a bulk isotropic phase showing boundary-induced local lamellar ordering (radius $R=5\xi$, $A=0.1$  (isotropic state), $C=2$, $\kappa^2 = 0.5$, anchoring: $\psi=e^{i\pi/2}$, $\vec{N}$ parallel to the boundary). 
    Exponential decay of lamellar order $\ps$ with distance from the surface $r$. 
    \textbf{(inset)} Snapshot of $\ps$ with layer normal shown in red. 
    \textbf{(b)} Inverse of the exponential decay length $\zeta$ as a function of Ginzburg parameter $\kappa$ for the isotropic phase confined between planar walls. 
    \textbf{(c)} Lamellar phase field in the vicinity of an inclusion (same parameters as \textbf{(a)} except $R=4\xi$ and $A=-1$ (lamellar state)).}
    \label{fig:col}
\end{figure*}

To explore the capacity of this model to describe smectic defects, consider a circular confining domain with
boundary conditions requiring a single $+1/2$ disclination (\fig{fig:defects}a). 
After minimization of the free energy, $\RE{\Psi}$ exhibits the lamellar structure around the disclination (\fig{fig:defects}a.i). 
Since $\vec{N}$ is the layer normal, it mirrors the layers (\fig{fig:defects}a.i-ii). 
The lamellar structure exhibits the expected symmetries of a $+1/2$ disclination and deformations are primarily bend on one side of the defect and splay on the other\cite{Zhang2018}. 
The lamellae are highly ordered away from the defect with $\ps\to\ps^\text{eq}$. However, $\ps\to0$ in the defect core (\fig{fig:defects}a.ii), verifying that $\tens{E}$-theory permits a finite sized defect core size. 
The deformations are principally curvature distortions, rather than compression, which is reflected in a constant phase everywhere in the vicinity of the disclination (\fig{fig:defects}a.iii). 
We find no evidence of any artificial order parameter melting where $\phi\to-\phi$, meaning that the non-physical free energy penalty observed in scalar theories\cite{Pevnyi2014} is circumvented. 
The situation is analogous for a $-1/2$ disclination (\fig{fig:defects}b): 
The layers are visualized by $\RE{\Psi}$ (\fig{fig:defects}b.i), with perpendicular layer normals (\fig{fig:defects}b.ii). 
The defect core is again seen to be locally disordered with no variation in phase, indicating negligible compression. 
In both $\pm1/2$ disclinations, the free energy density is largest in the immediate vicinity of the cores (\figsi{fig:defectsEnergy}). 
Not only is $f^\text{bulk}$ non-constant only at the core, but the deformation energy densities are strongly localized\cite{Zhang2018}. 

In addition to disclinations, lamellar states can support edge dislocations. 
While the phase $\phi$ is physically invariant to a global shift, it is set to vary linearly at the circular confining boundaries as $\phi=\theta/2$ for polar angle $\theta$ in \fig{fig:defects}c. 
This results in a dislocation: An extra layer is generated on the bottom half of \fig{fig:defects}c.i. 
While the lamellar order $\ps$ still decreases in the defect core (\fig{fig:defects}c.ii) and the order parameter variations are still localized around the core (\figsi{fig:defectsEnergy}), the phase changes by $\pi$ around the dislocation (\fig{fig:defects}c.iii). 
The occurrence of independent disclinations (\fig{fig:defects}a-b) and dislocations (\fig{fig:defects}c) highlights a strength of $\tens{E}$-theory: 
since theories of $\phi$ alone cannot model independent disclinations and models that simulate $\tens{Q}$ near the nematic-smectic transition cannot replicate dislocations. 
While disclinations and dislocations are considered separately in \fig{fig:defects}, they can co-reside in a single defect (\fig{fig:relax}). 

We now consider the role of defects in lamellar states evolving to equilibrium by simulating 2D systems with a deep quench from the isotropic to lamellar state, and periodic boundary conditions (\fig{fig:relax}). 
At first, the system is disordered (\fig{fig:relax}a), but relaxes through defect annihilation (\fig{fig:relax}b) to form many locally ordered domains (\fig{fig:relax}c). 
However, even at the longest times, the system remains disordered on macroscopic scales: 
It is kinetically arrested into a glassy configuration\cite{boyer2002} with a non-zero number of defects (\fig{fig:relax}d). 

To clarify this pinning of long-lived non-equilibrium structures, we compare simulations of large and small systems. 
While the small system routinely relaxes to the fully ordered lamellar state (\fig{fig:relax}e-inset and \movie{mov:fullOrder}) with $\lim_{t\to\infty}\ps \to \ps^\text{eq}$, the large system never reaches the global equilibrium (\fig{fig:relax}c). 
Correspondingly, the free energy of the small system rapidly approaches $f^\text{eq}$, the equilibrium defect free value; whereas, the large system is inevitably trapped away from equilibrium (\fig{fig:relax}e). 
Snapshots and associated videos show that both disclinations and edge dislocations are pinned\cite{Harrison2002} (\fig{fig:relax}f-g and \movies{mov:PSI}{mov:FreeEn}). 
While strikingly different than the relaxation dynamics of nematics\cite{liu1997,shendruk2015}, this highlights the importance of defects in lamellar ordering kinetics and the challenge posed for lamellar self-assembly\cite{Hur2015,kim2019,hur2018,rottler2020,Schneider2021}. 
The kinetic arrest of coarsening and long-lived domains are associated with pinned defects (\fig{fig:relax}e)\cite{boyer2002,hou1997} implies an energy barrier associated with the sliding of dislocations with respect to the lamellar structure. 
This non-zero Peierls-Nabarro energy barrier\cite{Pevnyi2014,hocking2021} indicates the validity of our formulation. 

The presence of inclusions embedded within the lamellar material can act to locally order layers or to induce additional defects. 
We evaluate boundary-induced lamellar ordering within an isotropic fluid ($A>0$), due to strong anchoring to a circular inclusion (\fig{fig:col}a). 
An inclusion with strong planar anchoring of $\vec{N}$ and $\psi=e^{i\pi/2}$ locally layers the smectic but the ordering rapidly decays (\fig{fig:col}a). 
By fitting an exponential to $\ps$ in a channel geomtery, we extract the decay length $\zeta$ (\fig{fig:col}b). 
We see that the decay length varies inversely with the Ginzburg parameter $\kappa$, indicating $\zeta$ varies linearly with lamellar coherence length $\xi$. 
This demonstrates the $\tens{E}$-formalism can be employed for nontrivial geometries. 
While strong anchoring locally orders the isotropic phase, it induces a pair of defects in the lamellar phase (\fig{fig:col}c; \movie{mov:colloid}). 
The topological charge of the circular inclusion is neutralized by the two $-1/2$ disclinations on opposite poles of the inclusion. 
Outside of the defect cores, the smectic remains well ordered and the deformation free energy contributions are localized around the inclusion (\figsi{fig:colloidEnergy}).

We have proposed a complex, symmetric, traceless, globally gauge invariant, normal, tensorial order parameter $\tens{E}\left(\vec{r}, t\right)$ for describing simple smectic phases at large scales. 
As a second-rank tensor, $\tens{E}$ encodes the apolar nature of the layer normal in an arbitrary reference frame and resolves the ambiguity of using the scalar phase alone. 
It avoids employing a microscopic approach, such as density functional theory\cite{wittmann2021} or particle-based simulations\cite{Monderkamp2021,Monderkamp2022}, which would also bypass such ambiguities at the cost of computationally expensive simulations. 
By conjoining local layer orientation and the extent of ordering into a single mathematical object, the $\tens{E}$ tensor can reproduce both disclination and dislocation defects with finite defect cores. 
While individual singularities can be analytically handled through local branch cuts, whether in $\vec{n}\to-\vec{n}$ for nematics or $\phi\to-\phi$ for smectics, the tensor order parameter description globally eliminates this ambiguity in a numerically pragmatic manner. 
Akin to the nematic $\tens{Q}$ tensor, this has the numerical advantage of avoiding point singularities. 
Though we restricted consideration to the simplest lamellar systems, generalizing to more complex smectics, including smectic-A or -C through coupling to $\tens{Q}$-theories, is conceptually straightforward\cite{mukherjee2001,das2008}. 
We expect this framework to be advantageous for simulating colloidal smectics\cite{cluzeau2001,pratibha2010,Honglawan2015,puschel2017,rasi2018,gharbi2018,do2020}, smectic-isotropic interfaces\cite{harth2009,vitral2019}, smectic-smectic emulsions\cite{radzihovsky2017}, smectics in contact with active material\cite{guillamat2017} and swimming bacteria in smectics\cite{ferreiro2018}. 

\section*{Acknowledgements}
We thank Linda S. Hirst 
for useful discussions. 
This research has received funding (TNS) from the European Research Council (ERC) under the European Union’s Horizon 2020 research and innovation programme (Grant agreement No. 851196) and we gratefully acknowledge funding from EPSRC.


\dummyapp{sctn:methods}

\dummyapp{sctn:lagrange}

\dummyapp{sctn:defects}

\dummyapp{sctn:viz}

\dummyapp{sctn:movies}
\dummymov{mov:fullOrder}
\dummymov{mov:PSI}
\dummymov{mov:PHI}
\dummymov{mov:REPSI}
\dummymov{mov:FreeEn}
\dummymov{mov:colloid}

\dummyapp{sctn:sifigs}
\dummyfig{fig:defectsEnergy}
\dummyfig{fig:colloidEnergy}

\bibliography{refSmectics}

\end{document}